\documentclass[lettersize,journal]{IEEEtran}

\usepackage{array}
\usepackage[hang]{footmisc}

\usepackage[subrefformat=parens,labelformat=parens,caption=false,font=footnotesize]{subfig}
\usepackage{amsmath, amssymb, amsthm}
\usepackage{cite,cleveref}

\usepackage[nolist,nohyperlinks]{acronym}
\begin{acronym}
    \acro{AP}{access point}
    \acro{ADAS}{advanced driver-assistance systems}
    \acro{CCDF}{complementary cumulative density function}
    \acro{CDF}{cumulative density function}
    \acro{CSMA}{carrier sense multiple access}
    \acro{CFA}{conditional false-alarm}
    \acro{CD}{conditional detection}
    \acro{PDF}{probability density function}
    \acro{PPP}{Poisson point process}
    \acro{BS}{base station}
    \acro{LOS}{line of sight}
    \acro{ROC}{receiver operating characteristic}
    \acro{SINR}{signal-to-interference-plus-noise ratio}
    \acro{SIR}{signal-to-interference ratio}
    \acro{SNR}{signal-to-noise ratio}
    \acro{mm-wave}{millimeter wave}
    \acro{PV}{Poisson-Voronoi}
    \acro{LOS}{line-of-sight}
    \acro{NLOS}{non line-of-sight}
    \acro{PGFL}{probability generating functional}
    \acro{PLCP}{Poisson line Cox process}
    \acro{MBS}{macro base station}
    \acro{MAC}{medium access control}
    \acro{PLT}{Poisson line tessellation}
    \acro{SBS}{small cell base station}
    \acro{RAT}{radio access technique}
    \acro{RCS}{radar cross section}
    \acro{5G}{fifth generation}
    \acro{PLP}{Poisson line process}
    \acro{UE}{user equipment}
    \acro{BPP}{binomial point process}
    \acro{BLP}{binomial line process}
    \acro{BLCP}{binomial line Cox process}
    \acro{RSU}{road-side unit}
    \acro{MD}{meta distribution}
    \acro{SF}{signal fraction}
    \acro{PCP}{Poisson cluster process}
    \acro{UCB}{Upper Confidence Bound}
    \acro{TS}{Thompson Sampling}
    \acro{CSP}{conditional success probability}
    \acro{MAB}{multi-armed bandit}
    \acro{QoS}{quality of service}
\end{acronym}

\usepackage{graphicx,wrapfig}

\newtheorem{theorem}{Theorem}[]
\newtheorem{corollary}{Corollary}[]

\newtheorem{lemma}[]{Lemma}
\newtheorem{remark}{Remark}[]

\crefname{figure}{fig.}{Fig.}
\crefname{section}{sec.}{Sec.}

\usepackage{algorithm}
\usepackage{algpseudocode}
\usepackage{scalerel}
\usepackage{multicol}
\usepackage{setspace}
\newtheorem{definition}{Definition}

\usepackage{bbm}
\usepackage[normalem]{ulem}
\usepackage{color}
\usepackage{dsfont}
\usepackage{bm}
\usepackage{setspace}

\usepackage{caption}
\usepackage{mathtools}
\begin{document}
\title{Distribution Bounds on the Conditional ROC in a Poisson Field of Interferers and Clutters}

	\author{
 Gourab Ghatak, {\it Member, IEEE}
 \thanks{The author is with the Department of Electrical Engineering at the Indian Institute of Technology (IIT) Delhi, New Delhi, India 110016. Email: gghatak@ee.iitd.ac.in.}
 }

\maketitle

\begin{abstract}
We present a novel analytical framework to characterize the distribution of the conditional receiver operating characteristic (ROC) in radar systems operating within a realization of a Poisson field of interferers and clutters. While conventional stochastic geometry-based studies focus on the distribution of signal to interference and noise ratio (SINR), they fail to capture the statistical variations in detection and false-alarm performance across different network realizations. By leveraging higher-order versions of the Campbell–Mecke theorem and tools from stochastic geometry, we derive closed-form expressions for the mean and variance of the conditional false-alarm probability, and provide tight upper bounds using Cantelli’s inequality. Additionally, we present a beta distribution approximation to capture the  meta-distribution of the noise and interference power, enabling fine-grained performance evaluation. The results are extended to analyze the conditional detection probability, albeit with simpler bounds. Our approach reveals a new approach to radar design and robust ROC selection, including percentile-level guarantees, which are essential for emerging high-reliability applications. The insights derived here advocate for designing radar detection thresholds and signal processing algorithms based not merely on mean false-alarm or detection probabilities, but on tail behavior and percentile guarantees.
\end{abstract}


\section{Introduction}
\subsection{Context and Related Work}
Radar systems form a core component of modern sensing and surveillance applications, ranging from automotive safety and air traffic control to autonomous navigation and battlefield awareness. The performance of such systems is conventionally evaluated using the \ac{ROC}, which characterizes the trade-off between the probability of detection and the probability of false-alarm. As radar technologies evolve toward more dense, dynamic, and shared spectral environments, understanding the variability in detection performance due to random interference and clutter becomes increasingly critical. In this regard, stochastic geometry has been recently employed to account for the stochastic nature of interferer and clutter locations to investigate their impact on radar performance~\cite{ram2020estimating, ram2022estimation}.

This modeling and characterization of interference experience by radar systems and its impact on the performance have been studied actively in the recent past, e.g., see~\cite{schipper2015simulative, al2017stochastic, munari2018stochastic, chu2020interference, wiame2024joint, weng2025frft}. In~\cite{brooker2007mutual}, the authors demonstrated significant degradation in detection due to the interference, while authors in~\cite{goppelt2011analytical} discussed the impact of ghost targets on frequency modulated continuous wave (FMCW) radars. Schipper {\it et al}~\cite{schipper2015simulative} analyzed radar interference using traffic flow patterns to model vehicle distribution along roadways.

In the context of stochastic geometry the authors in~\cite{ghatak2022radar, al2017stochastic} utilized a \ac{PPP} to model vehicular one-dimensional radar distribution on highways to evaluate mean \ac{SIR}. Similarly,~\cite{munari2018stochastic} employed the strongest interferer approximation method to analyze radar detection range and false alarm rates. In~\cite{fang2020stochastic}, radar detection probability is derived for targets with fluctuating radar cross-sections (RCS) modeled with Swerling-I and Chi-square frameworks. These two-lane based studies of automotive radars was extended to a multi-lane case in~\cite{chu2020interference} through a marked point process model. The work by \cite{ma2024meta} extended the fine-grained meta-distribution analysis to joint sensing and communication networks.

\subsection{Limitations of Current Studies}
{\bf Absence of stochastic \ac{ROC} characterization:} Most stochastic geometry based studies on radar systems focus on the distribution of the \ac{SINR}, sometimes inaccurately termed as the {\it detection probability}~\cite{ghatak2021fine} or the probability of range estimation~\cite{al2017stochastic}. In contrast, traditional radar system analysis considers the type-I and type-II errors, or the probability of detection and the probability of false-alarm, together referred to as the \ac{ROC}. An analytical framework to derive the distribution of the \ac{ROC} across different network realizations is currently unknown. This inhibits designers to effectively employ the insights from stochastic geometry based studies in their system design.

{\bf Absence of a fine-grained \ac{ROC} analysis:} More critically, even if the probabilities of detection and false-alarm are obtained, they do not provide any insight into the variance of the radar performance across different network realizations. For example, a false-alarm probability of 0.1 may either mean that all radars in the network experience a false-alarm rate of 10\%, or it may mean that 10\% of the radars always raise a false-alarm while the rest have perfect detection and no false-alarms. Without a fine-grained insight derived from the distribution of the conditional \ac{ROC}, quantitative guarantees on the percentile performance is infeasible. This is critical in future sensing technologies where key performance indicators include reliability and service guarantee.

\subsection{Contributions, Organization, and Notation}
We setup our framework in Section~\ref{sec:prel} and outline the metrics of interest, i.e., \ac{CFA} and \ac{CD} probabilities. Then, we make the following contributions.
\begin{enumerate}
    \item We define and derive the \ac{CFA} probability of radar detection in a Poisson field of interferers or clutter and show that it takes the form of a random sum-product functional. Since deriving the distribution of the same is infeasible, we derive a Cantelli-type bound for the same by leveraging the first and the second moments. Furthermore, based on the first two moments, we derive a beta-distribution based approximation for the distribution of \ac{CFA}. This is presented in Section~\ref{sec:CFA}.
    \item We extend the results of the \ac{CFA} probability to characterize the \ac{CD} probability that consists of an additional integral over the sum-product functional. This inhibits the derivation of the second moment, and hence we rely Markov inequality to derive the upper bound of the tail probability of the \ac{CD} probability. Nevertheless, for deterministic signal power, we derive the second moment enabling us to employ the Cantelli inequality for a more accurate bound. This is presented in Section~\ref{sec:CD}.
    \\
    {\bf Key novelty in 1) and 2):} Technically, unlike other meta-distribution based analyses especially that of \ac{SINR}, e.g., see \cite{haenggi2021meta, ghatak2021fine}, the distribution of \ac{CFA} probability is intractable due to the structure of its characteristic function. This requires new analysis tools based on higher-order Campbell-Mecke formulae. To the best of our knowledge, this is the first work that derives a bound and approximation of the meta-distribution of the interference or clutter return and a bound on the total signal power observed by the radar.
    
    \item {\bf System design insights:} From a practical perspective, our framework implies that traditional radar-design principles, e.g., {\it constant false-alarm rate} threshold selections may no longer hold, precisely since they are based on the first moment of the \ac{CFA} probability. This necessitates a new approach to radar system design based on confidence regions of false-alarm rather than constant false-alarm rates. With the help of numerical examples, we demonstrate this novel system-design insight. This is presented in Section~\ref{sec:NRD}.
\end{enumerate}

{\bf Notation:} Capital letters represent random variables, e.g., $P_{{\rm FA}_\Phi}$ represents the conditional false-alarm probability. Scalars are presented by small letters, e.g., $p_{\rm FA}$ represents the false-alarm probability across different network instances. $\mathbb{E}[\cdot]$, $\mathbb{V}(\cdot)$ and $\mathbb{P}(\cdot)$ represent the expectation, variance, and probability operators, respectively. 

\section{System Model and Performance Metrics}
\label{sec:prel}
Consider an ego radar located at the origin of the $d-$dimensional Euclidean plane. For an appropriate choice of $d$, different radar settings can be emulated For example, $d = 1$ corresponds to an automotive radar with target and interferers located along a highway~\cite{ghatak2021fine,al2017stochastic}. Similarly, $d = 2,$ and $3$ can represent the case of planar and aerial targets. We consider a generic model where the mean reflected signal power from the target conditioned on its distance from the radar is represented by $w_0$ (this includes the impact of the transmit and receive antenna gains, path loss, and transmit power). In addition to the average mean power, the target returns suffer from a fluctuating radar cross section $\sigma_0^2$. Thus the received signal power is $S = \sigma_0^2w_0$. The distribution of $\sigma_0^2$ is assumed to be generic for the development of our framework, while we derive simpler expressions for some special cases, e.g., the Swerling-I model.

\subsection{Unified Model for Clutter Returns and Interference} Consider a spatial stochastic process of nodes on $\mathbb{R}^d$ that can either emulate clutters, or interferers, or both. For traditional radar performance studies, clutter returns are considered as a key parameter since they create ghost targets that lead to a higher false-alarm rates. On the contrary, in case of automotive radars, while detecting the target vehicle, the radar observes reflections from the clutter resulting in ghost targets, which deteriorates the estimation efficiency. In particular, in the EU project MOSARIM~\cite{kunert2012eu}, it is experimentally discovered that the signals from interfering vehicles are unlikely to cause ghost targets; rather, they create noise-like combined interference. Let the locations of these nodes be modeled as points of a homogeneous \ac{PPP} $\Phi$ with intensity $\lambda$.
\begin{remark}
In this study, we refer to the points of $\Phi$ interferers, however, we emphasize that the consideration of $\Phi$ as a clutter set necessitates only a modification of the corresponding path-loss model.    
\end{remark}
Denote the mean of the interfering signals as $w_i, i \in \Phi$. In addition, the interfering signals suffer from independent and identically distributed fast-fading $h_i, i \in \Phi$. For tractability, we assume the fast fading to be Rayleigh distributed. It must be noted that extensions to other models such as Nagagami-$m$ is straightforward, albeit leading to more complicated expressions without leading to additional insights. Thus the total interference power received at the radar is $I = \sum_\Phi w_ih_i$. Finally, let us denote the noise power as $N_0$.

\subsection{Metrics of Interest}
\begin{figure*}
    \centering
    \includegraphics[width=0.84\textwidth, height=0.33\textwidth]{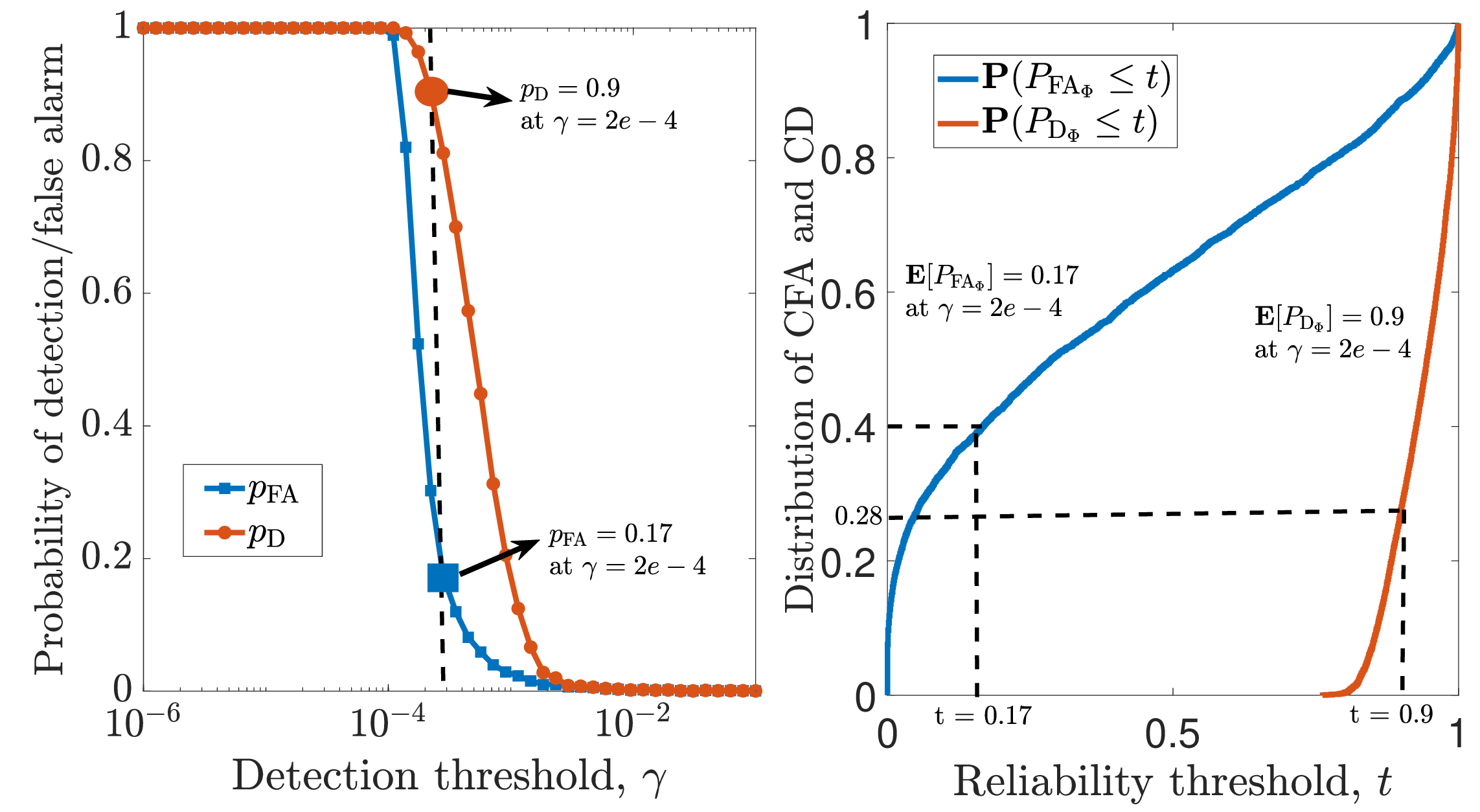}
    \caption{Illustration of the insights from fine-grained (right) analysis as compared to the standard ROC (left).}
    \label{fig:illustration}
\end{figure*}

We define the detection event as $S + I + N_0 = w_0\sigma_0^2 + \sum_\Phi w_ih_i + N_0 \geq \gamma$ and the false-alarm event as $I + N = \sum_\Phi w_ih_i + N_0 \geq \gamma$, for an appropriate detection threshold $\gamma$.
\begin{definition}
    The conditional detection probability and the conditional false-alarm probabilities are respectively defined as $P_{{\rm D}_\Phi} \coloneq \mathbb{P}\left( w_0\sigma_0^2 + \sum_\Phi w_ih_i + N_0 \geq \gamma \mid \Phi\right)$ and $P_{{\rm FA}_\Phi}\coloneq \mathbb{P}\left( \sum_\Phi w_ih_i + N_0 \geq \gamma \mid \Phi\right)$.
\end{definition}
Unlike the probability of detection $(p_{\rm D})$ and the probability of false-alarm $(p_{\rm FA})$ that are constants for a given $\gamma$, the \ac{CD} probability and \ac{CFA} probability are random variables. It is precisely these random variables that we analytically characterize in this paper. Indeed the means of these random variables are $p_{\rm D}$ and $p_{\rm FA}$, respectively. This is further illustrated in Fig.~\ref{fig:illustration} where on the left, we plot the components of the ROC, i.e., $p_{\rm D}$ and $p_{\rm FA}$ with respect to $\gamma$. As an example, for $\gamma = 2e-4$, i.e., $\approx -36$ dBW, we note that $p_{\rm FA}$ is 0.17 while $p_{\rm D}$ is 0.9. However, these numbers do not reveal the variance of performance across the network, and is only revealed by the figure on the right hand side where we plot the distribution of \ac{CFA} and \ac{CD} for $\gamma = 2e-4$. This reveals that although $p_{\rm FA} = 0.17$, over 60\% of the radars experience a higher false-alarm rate than 0.17. Similarly, although $p_{\rm D} = 0.9$, 28\% of the radars in the network experience a poorer detection probability 0.9.

In order to analytically investigate the distribution of the \ac{CD} and \ac{CFA} probabilities, we need to characterize higher order moments of the point processes. One such result is the second order Campbell-Mecke theorem stated below.

\begin{lemma}~\cite{chiu2013stochastic}
Let $\Phi$ be a simple and locally finite point process on $\mathbb{R}^d$, and let 
$f: \mathbb{R}^d \times \mathbb{R}^d \times \mathcal{N} \to [0, \infty)$ be a measurable function, 
where $\mathcal{N}$ denotes the space of locally finite counting measures on $\mathbb{R}^d$. 
Then the following identity holds:
\begin{align}
&\mathbb{E} \left[ \sum_{x \in \Phi} \sum_{\substack{y \in \Phi \\ y \neq x}} f(x, y, \Phi) \right] \nonumber \\
&= \iint_{\mathbb{R}^d \times \mathbb{R}^d} \mathbb{E} \left[ f(x, y, \Phi \cup \{x, y\}) \right] 
\rho^{(2)}(x, y) \, dx \, dy,
\end{align}
where $\rho^{(2)}(x, y)$ denotes the second-order product density of $\Phi$. The function $f$ 
may depend on the configuration $\Phi$, and the expectation on the right-hand side is taken with 
respect to the law of $\Phi$. In addition, if $\Phi$ is a stationary \ac{PPP} with intensity $\lambda > 0$, then 
$\rho^{(2)}(x, y) = \lambda^2$, and the identity simplifies to:
\begin{align}
\mathbb{E} \left[ \sum_{x \in \Phi} \sum_{\substack{y \in \Phi \\ y \neq x}} f(x, y) \right]
= \lambda^2 \iint_{\mathbb{R}^d \times \mathbb{R}^d} f(x, y) \, dx \, dy.
\end{align}
\end{lemma}

\section{Distribution Bound for the \ac{CFA} Probability}
\label{sec:CFA}
In this section we derive bounds and an accurate approximation for the distribution of the \ac{CFA} probability. The random \ac{CFA} probability is derived as
\begin{align}
    P_{\rm FA_\Phi} &=  \mathbb{P}\left(\sum_{i \in \Phi} w_i h_i + N_0 \geq \gamma \mid \Phi\right) \nonumber \\
    &=\mathbb{P}\left(\sum_{i \in \Phi} w_i h_i \geq \gamma  - N_0 \mid \Phi\right) \nonumber \\
    &= \sum_{i \in \Phi}\left(\prod_{j \in \Phi\backslash i} \frac{r_j^\alpha}{r_j^\alpha - r_i ^\alpha}\right)\exp\left(-\frac{\gamma - N_0}{KPr_i^{-\alpha}}\right).
\end{align}
The above follows from the fact that since $\{h_i\}$ are iid exponentially distributed with parameter 1, the addition of the weights $w_i$ results in $\sum w_ih_i$ being a hypoexponential random variable with phase-type distribution with the above \ac{CCDF}~\cite{o1990characterization}. Note that the first term inside the summation, i.e., the product term does not contain the transmit power $P$ as a constituent since we assume that the power to be the same across all nodes. To simplify the notation, let us denote
\begin{align}
    f(x) = \exp\left(-\frac{\gamma - N_0}{KPx^{-\alpha}}\right); \qquad g(x,y) = \frac{y^\alpha}{x^\alpha - y^\alpha}. \label{eq:fg_approximation}
\end{align}
Accordingly, the \ac{CFA} probability is a special form of sum-product functional:
\begin{align}
F = \sum_{x \in \Phi} f(x) \prod_{\substack{y \in \Phi \\ y \neq x}} g(x, y),
\end{align}
where \( f: \mathbb{R}^d \to [0, \infty) \) and \( g: \mathbb{R}^d \times \mathbb{R}^d \to [0, \infty) \) are measurable functions. Contrary to other stochastic geometry based analysis that characterize the distribution of the \ac{SINR}, we cannot employ inversion theorems (e.g., Gil-Pelaez theorem) to the quantity above. This is precisely due to the fact that the characteristic function of the \ac{CFA} probability is intractable. In this regard, we rely on higher order Campbell-Mecke formulae to derive bounds of the distribution of $P_{{\rm FA}_\Phi}$.
\begin{theorem}
    The mean of the \ac{CFA} probability is the probability of false-alarm across all network realizations, and is given by
    \begin{align}
       p_{\rm FA}= \lambda\int_{\mathbb{R}^d} \exp\left(-\frac{\gamma - N_0}{KPx^{-\alpha}}\right) \exp\left(-\lambda \int_{\mathbb{R}^d} \frac{x^\alpha}{y^\alpha - x ^\alpha} {\rm d} y\right) {\rm d}x.
    \end{align}
    Furthermore, the variance of the \ac{CFA} probability across different network realizations is
    \begin{align}
        \mathbb{V}(P_{{\rm FA}_\Phi}) = T_1 + T_2 - p_{\rm FA}^2,
    \end{align}
    where 
    \begin{align}
        T_1 &= \lambda\int_{\mathbb{R}^d} f^2(x)\exp\left(-\lambda\int_{\mathbb{R}^d}\left(1 - g^2(x,y)\right){\rm d}y\right){\rm d}x \nonumber \\
    T_2 &= \lambda^2\iint f(x)f(x')g(x,x')g(x',x) \cdot\nonumber \\
    & \exp\left(-\lambda \right. \int_{\mathbb{R}^d} 1 -\left.g(x,y)g(x',y) {\rm d}y{\rm d}x{\rm d}x'\right) \nonumber 
    \end{align}
    and the functions $f(\cdot)$ and $g(\cdot,\cdot)$ are defined in \eqref{eq:fg_approximation}.
    
\end{theorem}
\begin{IEEEproof}
The mean of $P_{{\rm FA}_\Phi}$ is evaluated as
    \begin{align}
       &p_{\rm FA} =  \mathbb{E}\left[\sum_{i \in \Phi}\left(\prod_{j \in \Phi\backslash i} \frac{r_j^\alpha}{r_j^\alpha - r_i ^\alpha}\right)\exp\left(-\frac{\gamma - N_0}{KPr_i^{-\alpha}}\right)\right] \nonumber \\
       & \overset{(a)}{=}\lambda \int_{\mathbb{R}^d} \mathbb{E}^{!x}\left[\exp\left(-\frac{\gamma - N_0}{KPx^{-\alpha}}\right)\prod_{j \in \Phi\backslash i} \frac{r_j^\alpha}{r_j^\alpha - x ^\alpha}\right] {\rm d}x \nonumber \\
       & \overset{(b)}{=} \lambda \int_{\mathbb{R}^d} \exp\left(-\frac{\gamma - N_0}{KPx^{-\alpha}}\right) \mathbb{E}\left[\prod_{j \in \Phi\backslash i} \frac{r_j^\alpha}{r_j^\alpha - x ^\alpha}\right] {\rm d}x \nonumber \\
       & \overset{(c)}{=}\lambda\int_{\mathbb{R}^d} \exp\left(-\frac{\gamma - N_0}{KPx^{-\alpha}}\right) \exp\left(-\lambda \int_{\mathbb{R}^d} \frac{x^\alpha}{y^\alpha - x ^\alpha} {\rm d} y\right) {\rm d}x.
       \label{eq:EF}
    \end{align}
The step $(a)$ follows from Campbell's theorem of homogeneous \ac{PPP}, i.e., $\mathbb{E}_{x\in \Phi}\left[f(x)\right] = \lambda\int_{\mathbb{R}^d} f(x)dx$. However the term inside consists of the product over the process $\Phi\backslash\{x\}$ which is taken care by considering the reduced Palm expectation $\mathbb{E}^{!x}[\cdot]$. The reduced Palm expectation is a tool of Palm calculus that evaluates the expectation by conditioning the existence of a point of $\Phi$ at $x$ and then removing it to take the expectation. Thanks to Slivnyak's theorem this is equal in probability law to the distribution of $\Phi$, as shown in step $(b)$. Finally, the step $(c)$ follows from the \ac{PGFL} of a \ac{PPP}.

Next, we turn our focus on the variance of $P_{{\rm FA}_\Phi}$ defined as
\begin{align}
    &\mathbb{V}(P_{{\rm FA}_\Phi}) = \mathbb{E}\left[P^2_{{\rm FA}_\Phi}\right] - \left(\mathbb{E}\left[P_{{\rm FA}_\Phi}\right]\right)^2. 
    \label{eq:var}
\end{align}
The second moment can be expressed as
\begin{align}
    P^2_{{\rm FA}_\Phi} &= \left(\sum_{i \in \Phi}\left(\prod_{j \in \Phi\backslash i} \frac{r_j^\alpha}{r_j^\alpha - r_i ^\alpha}\right)\exp\left(-\frac{\gamma - N_0}{KPr_i^{-\alpha}}\right)\right)^2 \nonumber \\
    &= \sum_{i \in \Phi}\sum_{k \in \Phi} \exp\left(-\frac{\gamma - N_0}{KPr_i^{-\alpha}}\right)\exp\left(-\frac{\gamma - N_0}{KPr_k^{-\alpha}}\right)  \nonumber \\
    &\qquad \qquad \prod_{j \in \Phi\backslash i} \frac{r_j^\alpha}{r_j^\alpha - r_i ^\alpha} \prod_{l \in \Phi\backslash k} \frac{r_l^\alpha}{r_l^\alpha - r_k ^\alpha}.
\end{align}
The above includes both the diagonal terms, i.e., $i = k$ and the off-diagonal terms, i.e., $i \neq k$. Let us now consider the notation introduced in \eqref{eq:fg_approximation}, and apply the second order Campbell-Mecke theorem to get
\begin{align}
    \mathbb{E}\left[P^2_{{\rm FA}_\Phi}\right]   = \lambda \int_{\mathbb{R}^d} \mathbb{E}\left[f^2(x)\prod g^2(x,y)\right] {\rm d}x +\lambda^2\iint_{\mathbb{R}^d \times\mathbb{R}^d}  \nonumber \\
    \mathbb{E}\left[f(x)f(x') \prod_{y \in \Phi}g(x,y) g(x',y)g(x,x')g(x',x)\right] {\rm d}x {\rm d}x'.
\end{align}
The first term is the contribution for $i = k$ while the second term is for $i \neq k$. Using reduced Palm expectations at two points (which for \ac{PPP} is again the same as the original law), we get
\begin{align}
   \mathbb{E}\left[P^2_{{\rm FA}_\Phi}\right]  &= \lambda\int_{\mathbb{R}^d} f^2(x)\exp\left(-\lambda\int_{\mathbb{R}^d}\left(1 - g^2(x,y)\right){\rm d}y\right){\rm d}x \nonumber \\
    &+\lambda^2\iint f(x)f(x')g(x,x')g(x',x) \cdot\nonumber \\
    & \exp\left(-\lambda \right. \int_{\mathbb{R}^d} 1 -\left.g(x,y)g(x',y) {\rm d}y{\rm d}x{\rm d}x'\right) 
    \label{eq:EF_square}
\end{align}
Substituting \eqref{eq:EF_square} and \eqref{eq:EF} in \eqref{eq:var} completes the proof.
\end{IEEEproof}

Leveraging the mean and the variance of the \ac{CFA} probability, we derive a bound on its \ac{CCDF} that is much tighter than the Chebyshev bound.
\begin{theorem}
    Using Cantelli's bound,
    \begin{align}
        \mathbb{P}\left(P_{\rm{FA}_\Phi} \geq t\right) \leq \frac{\mathbb{V}(P_{\rm{FA}_\Phi})}{\mathbb{V}(P_{\rm{FA}_\Phi}) + \left(t - p_{\rm FA}\right)^2}, \quad t \geq p_{\rm FA}.
    \end{align}
    \label{theo:Cant_FA}
\end{theorem}

\begin{remark}
    A relatively simple but loose bound follows from Markov's inequality
    \begin{align}
        \mathbb{P}\left(P_{\rm{FA}_\Phi} \geq t\right) \leq \frac{p_{\rm FA}}{t}.
    \end{align}
    Since we are mostly interested in the tail bounds of $P_{{\rm FA}_\Phi}$ for $t \geq p_{\rm FA}$ in order to derive performance guarantees on the radar performance, the above results in a valid probability bound (i.e., the right hand side is lower than 1).
\end{remark}

\subsection{Approximation using Beta Functions}
Next we provide a simple yet accurate approximation of the distribution of $P_{{\rm FA}_\Phi}$ using the beta distribution. This approximation arises naturally as the support of $P_{{\rm FA}_\Phi}$ is $[0, 1]$. The corresponding distribution can be approximated as the beta distribution by matching the first moment $M_1$ and the second moment $M_2$. Specifically, the MD can be approximately expressed as
\begin{align*}
    \bar{F}(\beta, t) \approx 1 - I_{\varepsilon}\left(\frac{k_1 k_2}{1-k_1}, k_2\right),
\end{align*}
where $I_{\varepsilon}(x,y) = \int_0^{1-\varepsilon} z^{x-1}(1-z)^{y-1}\mathrm{d}z/B(x,y)$ is the regularized incomplete beta function with $B(\cdot, \cdot)$ the beta function, $k_1 = P_{\rm FA}$, and 
$$k_2 = \left(p_{\rm FA} - \mathbb{E}\left[P^2_{{\rm FA}_\Phi}\right]\right)\left(1-p_{\rm FA}\right)/\left(\mathbb{E}\left[P^2_{{\rm FA}_\Phi}\right] - p_{\rm FA}^2\right).$$ The accuracy of the beta approximation is confirmed in Fig.~\ref{fig:bound}.


\section{Distribution Bound for the \ac{CD} Probability}
\label{sec:CD}
The critical difference between the \ac{CFA} and the conditional detection events is the added randomness due to the signal. The \ac{CD} probability is obtained as
\begin{align}
    &P_{\rm D_\Phi}  =\mathbb{P}\left(\sum_{i \in \Phi} w_i h_i \geq \gamma  - N_0 - w_0\sigma_0^2 \mid \Phi\right) \nonumber \\
    &= \int_0^{\gamma - N_0} \sum_{i \in \Phi}\left(\prod_{j \in \Phi\backslash i} \frac{r_j^\alpha}{r_j^\alpha - r_i ^\alpha}\right) \nonumber \\
    & \exp\left(-\frac{\gamma - N_0 -x}{KPr_i^{-\alpha}}\right) {\rm d}S(x)+ \mathbb{P}\left(S \geq \gamma - N_0\right) \nonumber \\
\end{align}
Thanks to the independence of the fluctuating target cross section and the interferer locations, the probability of detection across all network realizations is obtained by interchanging the expectation with respect to $\Phi$ and the integral with respect to $S$ by employing Fubini's theorem~\cite{veraar2012stochastic}:
\begin{align}
    &p_{\rm D} = \lambda\int_0^{\gamma -N_0}\int_{\mathbb{R}^d} \exp\left(-\frac{\gamma - N_0 - z}{KPx^{-\alpha}}\right) \nonumber \\
    &\exp\left(-\lambda \int_{\mathbb{R}^d} \frac{x^\alpha}{y^\alpha - x ^\alpha} {\rm d} y\right) {\rm d}x {\rm d}S(z)  +  \mathbb{P}\left(S \geq \gamma - N_0\right) \nonumber \\
    = &\lambda\int_0^{\gamma -N_0}\int_{\mathbb{R}^d} \exp\left(-\frac{\gamma - N_0 - z}{KPx^{-\alpha}}\right) \nonumber \\
    &\exp\left(-\lambda \int_{\mathbb{R}^d} \frac{x^\alpha}{y^\alpha - x ^\alpha} {\rm d} y\right) {\rm d}x {\rm d}\sigma_0^2(z/w_0)  +  \mathbb{P}\left(S \geq \gamma - N_0\right). \nonumber
\end{align}

\begin{corollary}
For the Swerling-I model for the target cross section, $\sigma_0^2$ is exponentially distributed, and accordingly, the detection probability has the following form.
\begin{align}
    &p_{\rm D} =\underbrace{\exp\left(-\frac{\gamma - N_0}{w_0\bar{\sigma}^2}\right)}_{A} + \frac{\lambda}{w_0\bar{\sigma}^2}\int_0^{\gamma -N_0}\int_{\mathbb{R}^d}\underbrace{\exp\left(-\frac{z}{\bar{\sigma}^2}\right)}_{B} \cdot \nonumber \\
    & \underbrace{\exp\left(-\frac{\gamma - N_0 - w_0z}{KPx^{-\alpha}}\right)}_{C} \underbrace{\exp\left(-\lambda \int_{\mathbb{R}^d} \frac{x^\alpha}{y^\alpha - x ^\alpha} {\rm d} y\right)}_{D} {\rm d}x  {\rm d} z . \nonumber
\end{align}
In the above the term $A$ corresponds to the event that the signal power alone is sufficient for crossing the detection threshold $\gamma$. The term $B$ corresponds to the contribution of the signal power in the total power when the former is below the detection threshold. The term $C$ accounts for the noise power and the term $D$ accounts for the contribution of the interference signals.
\end{corollary}

A Cantelli-type bound is challenging for the detection probability since the \ac{CD} probability contains two-terms where the first term itself is an integral. This results in an intractable second moment in general. Nevertheless, from Markov's inequality, we have the following bound.
\begin{corollary}
The upper bound on the distribution of the \ac{CD} probability is
        $\mathbb{P}\left(P_{\rm{D}_\Phi} \geq t\right) \leq \frac{p_{\rm D}}{t}$.
    \label{cor:Mark_D}
\end{corollary}

For the special case when the radar signal deterministic\footnote{This corresponds to the case when the impact of the fluctuating radar cross-section is averaged out using multiple measurements across different time slots or different channels.}, we can derive a Cantelli-type bound as discussed below.

\begin{lemma}
\label{lem:det_CD}
    In case the radar signal is deterministic, the second moment of the \ac{CD} probability can be derived as
    \begin{align}
        &\mathbb{E}\left[P^2_{{\rm D}_\Phi}\right]  = \nonumber \\
        &\begin{cases}
            &\lambda\int_{\mathbb{R}^d} f_0^2(x)\exp\left(-\lambda\int_{\mathbb{R}^d}\left(1 - g^2(x,y)\right){\rm d}y\right){\rm d}x \nonumber \\
        &+\lambda^2\iint f_0(x)f_0(x')g(x,x')g(x',x) \cdot\nonumber \\
        & \exp\left(-\lambda \right. \int_{\mathbb{R}^d} 1 -\left.g(x,y)g(x',y) {\rm d}y{\rm d}x{\rm d}x'\right); \; w_0\sigma_0^2 \leq \gamma - N_0. \\
        &1; \; w_0 \bar{\sigma}_0^2 > \gamma - N_0. 
        \end{cases}      
    \label{eq:ED_square}
    \end{align}
    where
    \begin{align}
        f_0(x) = \exp\left(-\frac{\gamma - N_0 - w_0 \bar{\sigma}_0^2}{KPx^{-\alpha}}\right), \nonumber 
    \end{align}
    and $g(x,y)$ is the same as the one given in \eqref{eq:fg_approximation}.
\end{lemma}

\section{Numerical Results and System Design Insights}
In this section, we discuss the accuracy of the bounds and the approximations derived in the previous sections and then outline system design insights based on them. We assume a transmit power $P_t = 10$ dBm and a path-loss exponent $\alpha = 2$. Furthermore, we assume a Swerling-1 model for the target cross section with mean 1 m.

\begin{figure}
    \centering
    \includegraphics[width=0.9\linewidth]{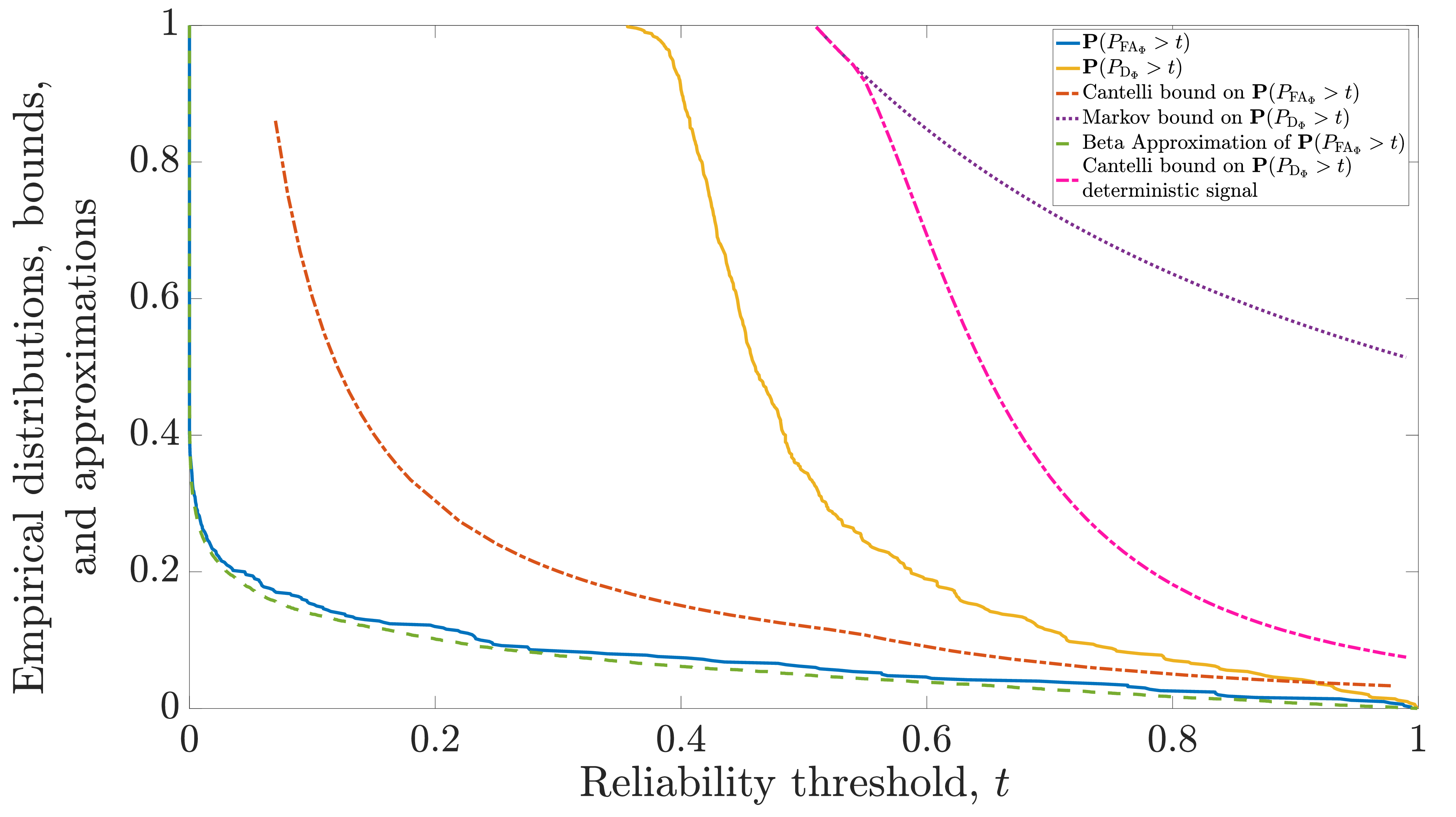}
    \caption{Empirical distributions of the \ac{CFA} and \ac{CD} probability along with the corresponding bounds and the beta-approximation for the distribution of the \ac{CFA} probability.}
    \label{fig:bound}
\end{figure}

First, Fig.~\ref{fig:bound} presents (with solid lines) the empirical distributions of the \ac{CFA} and the \ac{CD} probabilities for a fixed detection threshold $\gamma = 2e-4$, showcasing the variation of these quantities across realizations of a Poisson field of interferers. This reinforces the stochastic nature of radar performance metrics in interference-dominated environments. The line representing the beta distribution for the distribution of the \ac{CFA} probability closely tracks the empirical distribution obtained via Monte-Carlo simulations over multiple realizations of the Poisson interferer field. Importantly, the match is tight in the central region and the lower tail as well, which are critical for estimating high-reliability performance bounds. By using this beta fit, system designers can circumvent the need for extensive Monte Carlo simulations to estimate tail probabilities.

We further plot the derived bounds and show that due to the more sophisticated Canteli-type inequality employed for the \ac{CFA} (see Theorem~\ref{theo:Cant_FA}), the bound becomes tighter towards the right-tail. Recall that the bounds hold for the regions higher than their corresponding expected values. On the contrary, for the \ac{CD} probability Markov bound (Corollary~\ref{cor:Mark_D}) is rather loose. Nevertheless, as discussed in Lemma~\ref{lem:det_CD}, in case the deterministic part of the useful signal is estimated at the radar, a more useful Cantelli-type bound can be obtained as shown.

\label{sec:NRD}

\begin{figure}
    \centering
    \includegraphics[width=0.8\linewidth]{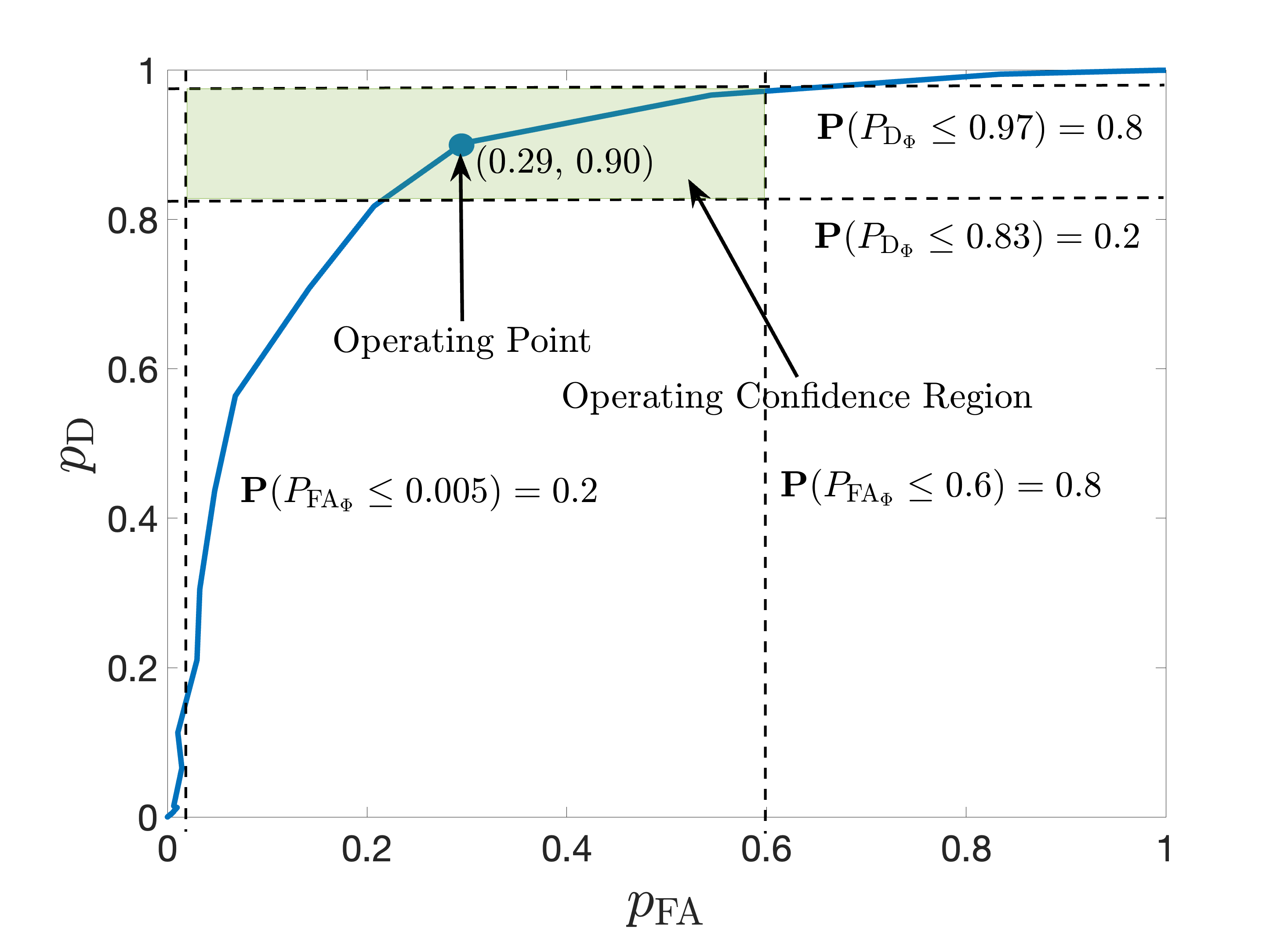}
    \caption{Illustration of a sample ROC empirically obtained while the bound based analysis reveals the operating confidence region.}
    \label{fig:ROC}
\end{figure}

Next, in Fig.~\ref{fig:ROC} we show how our analysis can be employed for developing a new fine-grained approach towards radar design. We show a sample ROC with an example operating point with $p_{\rm FA} = 0.29$ and $p_{\rm D} = 0.9$. Traditionally, radar engineers rely on constant false-alarm rate designs. However, from this figure we see that for the given operating point, although the mean false-alarm probability is 0.29, only 80\% of the radars in the network experience a false-alarm rate of less than 0.6. On the other extreme, we see that the best 20\% of the radars in the network achieve a false-alarm rate of 0.005. Indeed this is a high degree of variation in the performance, a fine-grained nuance that is missed by simply designing on the basis of $p_{\rm FA}$. Similarly, even though the mean detection probability is 0.9, we see that the designer can guarantee that 80\% of the radars will achieve a detection of more than 0.83 (i.e., the fraction of radars that have a conditional detection probability of less than 0.83 is 0.2). In fact, the top 20\% of the radars have a detection probability of over 0.97.

These observations are central to the motivation of our work -- moving from mean-value performance metrics to full distributions or meta-distributions. They underscore the importance of accounting for performance dispersion due to spatial randomness, which is especially relevant in automotive radar systems where reliability guarantees are essential. Thus, instead of designing to meet an average detection or false-alarm probability, engineers must ensure performance under worst-case or percentile constraints (e.g., 90th percentile CFA).

\section{Conclusions and Discussion}
We introduced a novel analytical framework to characterize the distributional behavior of the conditional receiver operating characteristic (ROC) for radar systems operating in stochastic environments modeled by a Poisson field of interferers and clutters. By shifting focus from conventional SINR-based metrics to the conditional false-alarm (CFA) and conditional detection (CD) probabilities, modeled as random variables, we captured the fundamental spatial variability in radar performance that is often overlooked in traditional analyses.

Using tools from stochastic geometry—including higher-order Campbell–Mecke theorems—we derived closed-form expressions for the first and second moments of the CFA probability and proposed Cantelli-type inequalities to bound its distribution. We also developed a beta distribution-based approximation to capture the meta-distribution of CFA with high accuracy. For the CD probability, we provided analytical bounds under both generic and deterministic signal power models, revealing the inherent difficulty in characterizing its full distribution due to the additional signal randomness.

Our numerical evaluations reveal that reliance on average performance metrics can be misleading, as a significant fraction of network realizations deviate from the mean behavior. This underscores the need for distribution-aware radar system design, particularly in high-reliability applications such as automotive radar, aerial surveillance, autonomous navigation, and collaborative sensing networks in smart cities. 

\bibliography{refer.bib}
\bibliographystyle{IEEEtran}

\end{document}